# The melting curve of Sodium: a theoretical approach

Jozsef Garai

*Department of Mechanical and Materials Engineering, Florida International University, Miami, USA*

*Electronic mail: jozsef.garai@fiu.edu*

New model describing the pressure effect on the melting temperature is proposed by using four assumptions.  One, the average wavelength of the phonon vibration at the Debye temperature corresponds to the length of the unit cell.  Two, the phonon vibration at the melting temperature is in self-resonance with the lattice vibration of the surface atomic/molecular layer.  Three, the phonon wavelength ratio of the Debye and the melting temperature does not be affected by the pressure.  Four the pressure reduces the anharmonic part of the vibration.  The relevant equations are derived and tested against the experiments of sodium with positive result.

Alkali metals have a simple structure and exhibit nearly free electron behavior at low pressure; however, the behavior of these metals become quite complex at high pressure[1-6].  The same behavior pattern is true for melting.  At low pressure the reported melting curves of Bridgman[7] fit well to the monotonically increasing Simon equation[8]:

$$\frac{p_m - p_0}{a} = \left(\frac{T_m}{T_0}\right)^c - 1 \qquad (1)$$

where $T_0$ and $p_0$ are coordinates of the triple point and a and c are constants characteristics of the substance.  At higher pressures Eq. (1) does not fit to experiments because the melting curves exhibit a maximum.  The maximum for Rubidium[9] is at 5 GP and 553 K.  Cesium exhibits two maximums[10], Cs I at 2.25 GPa and 464 K and Cs II at 4.72 GPa and 363 K.  The melting curve of Sodium shows similar characteristics[11,12] and exhibit a maximum in the bbc phase at 31 GPa and 1000 K.  At higher pressures the curve steeply decreases after the maximum.

In order to describe this unusual behavior of the melting curve of alkali metals Kechin[13] proposed to add a dumping function to the Simon equation as:

$$T_m = F(p)D(p). \qquad (2)$$

where F(p) is the Simon equation for the rising part and D(p) is a dumping function.  The equation is then:

$$T_m = T_0\left(1 + \frac{p_m - p_0}{a}\right)^b e^{-c(p_m - p_0)}. \qquad (3)$$

where a, b, and c are constants characteristics of the substance.  The equation reproduces the melting curves of halides but the good fitting requires adjusting three parameters which are characteristics of the substance. Ab initio molecular dynamics calculations have also been used successfully to reproduce the trend of the curve[14] and the melting curve[15]. The standard ab initio approach[ex. 16, 17] is complicated and expensive. In order to overcome the complexities, simple criterions are suggested here to define the pressure effect on the melting temperature. Using these criterions the relevant equations are derived and tested against the experimental melting curve of sodium.

It has been shown that at the Debye temperature the wavelength of the phonon vibration is equivalent with the smallest atomic unit of the crystal structure, which is the unit cell[18]. In a recent physical description of melting[19] it has been suggested that at the melting temperature the thermal phonon vibration is in self-resonance with the lattice vibration of the surface atomic/molecular layer. Since the average wavelength of the phonon vibration [$\bar{\lambda}$] at both the Debye and the melting temperature relates to lattice parameters as:

$$\bar{\lambda}_{T_D} \Rightarrow a/b/c \quad \text{and} \quad \bar{\lambda}_{T_m} \Rightarrow \frac{1}{n}d \quad \text{where } n \in \mathbb{N}. \tag{4}$$

when pressure is applied then these lattice parameters should change their size in proportion and the ratio of the two wavelengths related to these lattice parameters should remain constant. Thus the ratio of the corresponding phonon wavelength should not be affected by the pressure and:

$$\frac{\bar{\lambda}_{T_D}}{\bar{\lambda}_{T_m}} = \frac{\bar{f}_{T_m}}{\bar{f}_{T_D}} = \text{const}. \tag{5}$$

where $\bar{f}$ is the average frequency, $T_D$ refers to the Debye temperature and $T_m$ to the melting temperature. The melting temperature can be calculated then as:

$$T_m(p) = \left(\frac{\bar{f}_{T_m}}{\bar{f}_{T_D}}\right) T_D(p). \tag{6}$$

Equation (6) has been tested on several substances[20]. The agreement between theory and experiments were not satisfactory in all cases. Investigating the disagreement between the model and the experiments it has been concluded that the deviation most likely occurs because the effect of the anharmonic vibration is neglected. At higher pressures the anharmonic part of the vibration is reduced. This effect is taken into consideration by adding a pressure dependent anharmonic term to Eq. (6) as:

$$T_m = \left(\frac{\bar{f}_{T_m}}{\bar{f}_{T_D}} + c_{anh}\frac{\alpha_1}{K_o}p\right) T_D. \tag{7}$$

where $c_{anh}$ is a constant, $\alpha_1$ is the temperature derivative of the volume coefficient of thermal expansion, and $K_o$ is the bulk modulus at zero pressure and temperature.

At temperatures higher than the Debye temperature the relationship between the phonon frequency and the temperature is given as:

$$hf = k_B T \tag{8}$$

where h is the Planck constant, and $k_B$ is the Boltzmann constant. The frequencies of the phonon vibrations can be calculated as:

$$f = \frac{v_B}{\lambda} \tag{9}$$

where $v_B$ is the bulk sound velocity which is approximated as:

$$v_B = \sqrt{\frac{K}{\rho}} = \sqrt{\frac{KV_{mol}}{M}} \tag{10}$$

where $\rho$ is the density, M is the molar mass and $V_{mol}$ is the molar volume. The molar volume can be directly calculated by using the EoS of Garai[21] as:

$$V_{mol} = V_{o,mol} e^{\frac{-p}{ap+bp^2+K_o} + \left(\alpha_o + cp + dp^2\right)T_{m,p} + \left(1 + \frac{cp+dp^2}{\alpha_o}\right)^f gT_{m,p}^2} \tag{11}$$

where subscript o refers to the initial value at zero pressure and temperature. Thus $V_{o,mol}$ is the initial molar volume, $K_o$ is the initial bulk modulus and $\alpha_o$ is the initial volume coefficient of thermal expansion. In Eq. (11) a is a linear, b is a quadratic term for the pressure dependence of the bulk modulus, c is linear and d is a quadratic term for the pressure dependence of the volume coefficient of thermal expansion and f and g are parameters describing the temperature dependence of the volume coefficient of thermal expansion. The theoretical explanations for Eq. (11) and the physics of the parameters are discussed in detail[21]. Assuming that the product of volume coefficient of thermal expansion and the bulk modulus is constant allows calculating the bulk modulus[22] as:

$$K = \left(K_o + K_o' p\right) e^{-(\alpha_o + \alpha_1 T)\delta T} \tag{12}$$

where $K_o'$ is the pressure derivative of the bulk modulus and $\delta$ is the Anderson- Grüneisen parameter, which defined as:

$$\delta_T \equiv \left(\frac{\partial \ln B_T}{\partial \ln V}\right)_p = -\frac{1}{\alpha_{V_p}} \left(\frac{\partial \ln B_T}{\partial T}\right)_p = -\frac{1}{\alpha_{V_p} B_T} \left(\frac{\partial B_T}{\partial T}\right)_p. \tag{13}$$

The parameters in Eq. (12) are taken from the universal p-V-T form of the Birch-Murnaghan EoS[23-26], which is given as:

$$p = \frac{3K_o e^{-(\alpha_o + \alpha_1 T)\delta T}}{2} \left[\left(\frac{V_o e^{(\alpha_o + \alpha_1 T)T}}{V}\right)^{\frac{7}{3}} - \left(\frac{V_o e^{(\alpha_o + \alpha_1 T)T}}{V}\right)^{\frac{5}{3}}\right]$$
$$\left\{1 + \frac{3}{4}\left(K_o' - 4\right)\left[\left(\frac{V_o e^{(\alpha_o + \alpha_1 T)T}}{V}\right)^{\frac{2}{3}} - 1\right]\right\} \tag{14}$$

Please note that $\alpha_1$ is equivalent with g in Eq. (11). The average wavelength of the phonon vibration relating to the Debye temperature is approximated as:

$$\bar{\lambda}_{T_D} = a; c = \sqrt[3]{\frac{n_a V_{mol}}{N_A}} \tag{15}$$

where $n_a$ is the number of atoms in the unit cell and $N_A$ is the Avogadro's number. Substituting Eqs. (9), (10) and (15) into Eq. (8) allows calculating the Debye temperature as:

$$T_D = h k_B^{-1} M^{-\frac{1}{2}} N_A^{\frac{1}{3}} n_a^{-\frac{1}{3}} V(p,T)_{mol}^{\frac{1}{6}} K(p,T)^{\frac{1}{2}} \tag{16}$$

Substituting Eqs. (11) and (12) into Eq. (16) allows calculating the Debye temperature at the given pressure and the melting temperature as:

$$T_{D(T_m,p)} = \frac{h}{k_B \sqrt{M}} \sqrt[3]{\frac{N_A}{n_a}} \frac{\left( nV_o^m e^{\frac{-p}{ap+bp^2+K_o}+\left(\alpha_o+cp+dp^2\right)T_{m,p}+\left(1+\frac{cp+dp^2}{\alpha_o}\right)^f gT_{m,p}^2} \right)^{\left(\frac{1}{6}\right)}}{\sqrt{(K_o+K'_o p) e^{-(\alpha_o+\alpha_1 T_{m,p})\delta T_{m,p}}}} \times \tag{17}$$

The melting temperature can be calculated by repeated substitution of Eq. (6) as:

$$T_{m,p} = \lim_{n \to \infty} f^n(T_{m,p}) \tag{18}$$

where

$$f^n(T_{m,p}) = \left(\frac{T_{m0}}{T_{D0}} + c_{anh} \frac{\alpha_1}{K_o} p\right) \frac{h}{k_B \sqrt{M}} \sqrt[3]{\frac{N_A}{n_a}} \times$$

$$\frac{\left( nV_o^m e^{\frac{-p}{ap+bp^2+K_o}+\left(\alpha_o+cp+dp^2\right)T_{m,p(n-1)}+\left(1+\frac{cp+dp^2}{\alpha_o}\right)^f gT_{m,p(n-1)}^2} \right)^{\left(\frac{1}{6}\right)}}{\sqrt{(K_o+K'_o p) e^{-(\alpha_o+\alpha_1 T_{m,p(n-1)})\delta T_{m,p(n-1)}}}} \times \tag{19}$$

$$n \in \mathbb{N}^* \quad \text{and} \quad T_{m(0)} = T_{m0}$$

In equation (19) $T_{m0}$ is the melting temperature and $T_{D0}$ is the Debye temperature at atmospheric pressure. The convergence is fast and 15-20 iteration is sufficient.

Using the available experiments of Na[27-30] the universal Birch-Murnaghan and Garai EoS were determined by unrestricted fitting. The parameters are given in Table I and II. Using $7.7 \times 10^5$ value for the anharmonic constant, and the parameters of the EoS of Garai the melting curve is calculated. It can be seen that (Fig. 1) equation (18) reproduces the experiments with high accuracy with a maximum at 30 GPa and 1020 K.

TABLE I. Parameters of the universal (P-V-T) Birch-Murnaghan EoS used for the calculations.

| EoS-B-M | $K_o$ [GPa] | $K_o'$ | $\alpha_o$ ×10$^{-5}$ | $\alpha_{1*}$ ×10$^{-9}$ | $\delta$ | $\overline{\lambda}_{T_D} / \overline{\lambda}_{T_m}$ |
|---|---|---|---|---|---|---|
| Na | 6.3 | 3.82 | 50.5 | -183.9 | 0.039 | 1.66 |

* $\alpha_1$ is equivalent with g in the EoS of Garai.

TABLE II. Parameters of the Garai EoS used for the calculations.

| EoS-G | $V_o$ [cm$^3$] | $K_o$ [GPa] | $\alpha_o$ [×10$^{-5}$ K$^{-1}$] | a | b [×10$^{-3}$] | c [×10$^{-7}$] | d | g [×10$^{-9}$] | f |
|---|---|---|---|---|---|---|---|---|---|
| Na | 20.85 | 7.73 | 49.24 | 1.046 | -3.749 | -51.39 | 0 | -165.3 | 12 |

* The parameter g is equivalent with $\alpha_1$ in the universal Birch-Murnaghan EoS.

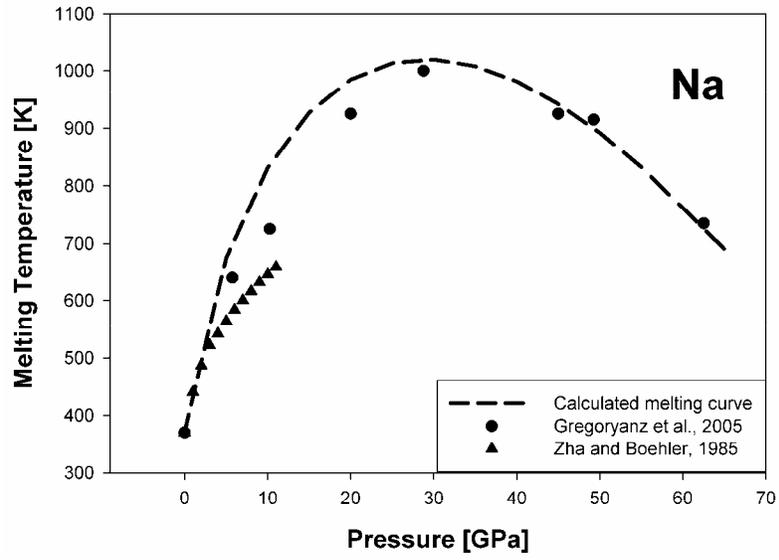

FIG. 1 The calculated pressure-melting temperature curve of Sodium is compared to experiments.